\documentstyle[epsf,prl,aps,multicol]{revtex}
\renewcommand{\narrowtext}{\begin{multicols}{2}
\global\columnwidth20.5pc\noindent}
\renewcommand{\widetext}{\end{multicols}
\global\columnwidth42.5pc}
\newcommand{\Ymark}
{{\normalsize\boldmath $Y\hspace{-1mm}ukawa~
I\hspace{-.3mm}n\hspace{-.3mm}stitute~K\hspace{-.5mm}yoto$}
\hfill {\rm\normalsize YITP-96-11}\\~ \\}
\newcommand{\sgn}{{\rm sgn}}
\newcommand{\tildk}{\widetilde k}
\newcommand{\tildm}{\widetilde m}
\newcommand{\ch}{\!\!\circ\!\!}
\begin{document}
\draft
\preprint{}
\title{\Ymark Spectral flow in the supersymmetric $t$-$J$ model 
with a $1/r^2$ interaction}
\author{ Takahiro Fukui \cite{Ema}} 
\address{Yukawa Institute for Theoretical Physics, 
Kyoto University, Kyoto 606-01, Japan}
\author{Norio Kawakami}
\address{Department of Applied Physics,
and Department of Material and Life Science,\\
Osaka University, Suita, Osaka 565, Japan}
\date{April 22, 1996}
\maketitle
\begin{abstract}
The spectral flow in the supersymmetric {\it t-J} model with $1/r^2$
interaction is studied by analyzing the exact spectrum 
with  twisted boundary conditions.
The spectral flows for the charge and spin sectors  
are shown to nicely fit in with  
the motif picture in the asymptotic Bethe
ansatz. Although fractional exclusion statistics
for the spin sector clearly
shows up in the period of the spectral flow  at half filling,
such a  property is generally hidden once
any number of holes are doped, because the commensurability 
condition in the motif is not met in the metallic phase.
\end{abstract}
\narrowtext
\section{Introduction}
Since the seminal work of Sutherland and Shastry\cite{SutSha,ShaSut},
the ground state properties in quantum systems
with adiabatic change of twisted boundary conditions
have attracted much current interest. 
Although the period of the full spectrum 
should be $2\pi$ as a function of the twist angle, 
each eigenstate has a period larger than $2\pi$ in general,
if we trace the flow of the initial state adiabatically.
This period provides us various information for quantum spin
systems, correlated electron systems, etc. 
For example, it can be a sensitive probe 
to examine whether particles form a bound state like Cooper
pair\cite{Sut,Aok}. 

One of the remarkable applications of this method
is to observe (fractional) exclusion statistics 
\cite{HalFES} directly in the spectral 
flow of the ground state for one-dimensional quantum systems
\cite{FukKaw2}.
For this purpose, it is crucial to sweep away all irrelevant 
interactions from the model, because the period
is quite sensitive to irrelevant interactions, 
leading to the period $4\pi$ if they exist
\cite{SutSha,FukKaw1}, and hiding the nature of exclusion statistics.
Among others, it is known that quantum many-body systems 
with $1/r^2$ interaction \cite{Hal,Sha} 
provide us with an ideal situation to observe  exclusion statistics,
i.e., these systems are completely free from 
irrelevant interactions. Motivated by this, we have recently 
studied the spectral flow in the 
Haldane-Shastry spin model with $1/r^2$ exchange
by generalizing the model 
to include twisted boundary conditions \cite{FukKaw2}.
We have indeed found that the period of the spectral flow
is controlled by the statistical interaction in
exclusion statistics.

In order to investigate exclusion statistics for
correlated electron systems, we wish  to ask a basic question, i.e.
what happens for the spectral flow when holes 
are doped into the Haldane-Shastry 
spin model. To address this question, 
we investigate in this paper the exact
spectral flow in the supersymmetric $t$-$J$ model
with $1/r^2$ hopping and exchange, which
was exactly solved for the first time by Kuramoto and Yokoyama with
periodic boundary conditions\cite{KurYok,HaHal,WLC}.
We extend the model to include the
twisted boundary conditions, and then discuss spectral properties 
of the model as a function of the twist angle.

The paper is organized as follows.
In the next section, we briefly mention how to 
generalize  the solvable supersymmetric {\it t-J}
model of $1/r^2$ interaction to be compatible 
with twisted boundary conditions.
In section III, we then derive the exact eigenstates
of the model by exploiting a Jastrow-type ansatz. 
In section IV, it is shown that 
the spectrum thus obtained can be given 
alternatively by the asymptotic Bethe ansatz (ABA).
It is then demonstrated that the spectral flow of the 
model is completely described in terms of the motif 
in the  ABA.  We find that once 
holes are doped into the Haldane-Shastry spin chain, 
the nature of fractional statistics can not be observed
in the period of the spectral flow, though
elementary excitations are still characterized by 
exclusion statistics. We claim that this is related to the 
breaking of the commensurability condition in the motif picture.
Section V is devoted to summary.

\section{model hamiltonian}

We consider the supersymmetric {\it t-J} model
with $1/r^2$ interaction \cite{KurYok}.
The model Hamiltonian is given  by 
\begin{eqnarray}
&H&={\cal P}\sum_{i\ne j}\sum_{l=-\infty}^{\infty}
\frac{1}{(i-j-lN)^2}\times\nonumber\\    
&&\left[-\sum_\sigma 
c_{i\sigma}^\dagger c_{j+lN\sigma}
+\left({\bf S}_i\cdot{\bf S}_{j+lN}
-\frac{1}{4}n_in_{j+lN}\right)\right]{\cal P},
\label{Ham1}
\end{eqnarray} 
where ${\cal P}$ is the projection operator to exclude the double
occupancies at each site,
$i$ and $j$ denote the indices for sites ($i,j=1,2,\cdots N$),
and $\sigma$ is a spin index labeled by $\sigma=\uparrow$ and
$\downarrow$. Note that
the summation with respect to $l$ is introduced to treat the system in 
a ring geometry \cite{Sut2}. 
Kuramoto and Yokoyama \cite{KurYok}
 have shown that this 
model can be solved exactly with periodic boundary conditions, 
in which the summation over $l$ results in the sine-inverse-square 
interaction.
This model, which includes the Haldane-Shastry
spin chain at half filling, provides us with a paradigm
of ideal exclusion statistics\cite{HalFES,KatKur}.

Preserving the integrability,
we wish to treat the model with twisted boundary conditions. 
Note that this type of generalization is not straightforward
for the present model because of the long-range 
nature of hopping and interaction.
As was the case for the spin chain\cite{FukKaw2}, 
it turns out that the summation over $l$ in eq.(\ref{Ham1}) 
after imposing twisted boundary conditions
plays a crucial role 
to preserve the integrability. 
The boundary conditions we impose in this paper are
\begin{eqnarray}
&&c_{j+lN\sigma}=
e^{-2\pi il\phi_\sigma}c_{j\sigma},\nonumber\\
&&c_{j+lN\sigma}^\dagger=
e^{2\pi il\phi_\sigma}c_{j\sigma}^\dagger.
\label{twistb}
\end{eqnarray}
In what follows, the angle $\phi_\sigma$ in unit of $2\pi$
defined above is referred to as the twist angle. 
Note that we can impose twisted boundary conditions 
independently on the spin and the charge degrees of freedom
with two kinds of twist angles.
To diagonalize the Hamiltonian, 
it is convenient to treat the system after 
a gauge transformation,
\begin{eqnarray}
&&c_{j\sigma}\rightarrow e^{-2\pi i\phi_\sigma j/N}
c_{j\sigma},\nonumber\\
&&c_{j\sigma}^\dagger\rightarrow e^{2\pi i\phi_\sigma j/N}
c_{j\sigma}^\dagger,
\label{GauTra}
\end{eqnarray}
as was done previously\cite{FukKaw2}.
Hence the transformed operators satisfy the periodic boundary 
condition.

In the remainder of the paper, we restrict ourselves to the case
of rational twist angles, 
\begin{equation}
\phi_\sigma =\frac{p_\sigma}{q_\sigma},
\label{FraAng}
\end{equation}
where $p_\sigma$ and $q_\sigma$ are integers.
This restriction enables us to carry out the summation over $l$ in
eq.(\ref{Ham1}), and the resultant Hamiltonian is well defined,
\begin{eqnarray}
H&=&\left(\frac{\pi}{N}\right)^2{\cal P}\sum_{i\ne j}
\Big[-\sum_\sigma J_{\phi_\sigma}(i-j)
c_{i\sigma}^\dagger c_{j\sigma}\nonumber\\
&+&J_{\phi_s}(i-j)S_i^+S_j^-
+J_{0}(i-j)\left( S_i^zS_j^z-\frac{1}{4}n_in_j\right)
\Big]{\cal P},
\label{Ham}
\end{eqnarray}
where $\phi_s=\phi_\uparrow-\phi_\downarrow$ and
the effective coupling constant is
\begin{equation}
J_\phi (n)\equiv\frac{1}{q^2}\sum_{m=0}^{q-1}
e^{2\pi ip(n+mN)/qN}\sin^{-2}\left[\frac{\pi (n+mN)}{qN}\right]
\label{EffCou}
\end{equation}
for $\phi=p/q$. The effect of twisted boundary conditions is now 
incorporated in the 
effective coupling (\ref{EffCou}), and we can solve (\ref{Ham}) with
periodic boundary conditions.
Putting $p=0$, we have $J_0(n)=\sin^{-2}\frac{\pi n}{N}$,
reproducing  the periodic model\cite{KurYok}.

In order to obtain the exact eigenstates of the Hamiltonian
(\ref{Ham}), we take the fully polarized state
$|P\rangle=|\uparrow\uparrow\cdots\uparrow\rangle$ 
as a reference state, and
introduce the fermionic hole operator $h_j$ by \cite{KurYok}
\begin{equation}
c_{i\uparrow}= h_i^\dagger,\quad 
c_{i\uparrow}^\dagger= h_i,
\end{equation}
and the  bosonic spin operator $b_j$ by
\begin{equation}
S_{i}^-=b_i^\dagger , \quad 
S_{i}^+=b_i, \quad
S_i^z=\frac{n_i}{2}-b_i^\dagger b_i,
\end{equation}
where $n_i=1-h_i^\dagger h_i$.
The Hamiltonian now reads
\begin{equation}
H=\left(\frac{\pi}{N}\right)^2
\left(T_\uparrow(\phi_\uparrow)+
T_\downarrow(\phi_\downarrow)+T_s(\phi_s) +H_{\rm int}+e\right),
\label{effham}
\end{equation}
where
\begin{eqnarray}
&&T_\uparrow(\phi_\uparrow) =
\sum_{i\ne j}J_{\phi_\uparrow}(i-j)h_j^\dagger h_i ,\nonumber\\
&&T_\downarrow(\phi_\downarrow) =
\sum_{i\ne j}J_{\phi_\downarrow}(i-j)
c_{j\downarrow}c_{i\downarrow}^\dagger,\nonumber\\
&&T_s(\phi_s)=\sum_{i\ne j}J_{\phi_s}(i-j)b_j^\dagger b_i ,\nonumber\\
&&H_{\rm int}=\sum_{i\ne j}J_0(i-j)
\left(n_i^{(b)}n_j^{(b)}+n_i^{(h)}n_j^{(b)}\right),\label{IntHam}\nonumber\\
&&e=-\frac{1}{3}M(N^2-1).
\end{eqnarray}
Here and in what follows, we omit ${\cal P}$ for simplicity,
though we regard the Hamiltonian as always accompanying this operator.

\section{exact solution for twisted boundaries}

In this section, we obtain the exact eigenstates
of (\ref{effham}) by exploiting a Jastrow-type ansatz for the 
corresponding wave functions\cite{Hal,Sha,KurYok}.  Let us first divide 
the total set of sites $\{1,2,,\cdots,N\}$ into three subsets 
$\{x_\alpha\}, \{y_l\}$ and $\{u_a\}$, which denote, respectively, the 
position of down-spins (bosons), holes (fermions) and up-spins
\cite{KurYok,WLC}.
We use Greek and Latin indices for bosons and fermions, respectively,
the numbers of which are denoted by $M$ and $Q$.
Then, we have $S^z=\frac{N-Q}{2}-M$.


As was found in \cite{KurYok}, electronic wave functions of
Jastrow type are the exact eigenfunctions for the {\it t-J} model
with periodic boundary conditions.
We can also expect such wave functions to be still 
eigenstates even for twisted boundaries (\ref{twistb}) if we treat
the twisted systems after the gauge transformation (\ref{GauTra}),
as was indeed demonstrated
for the Haldane-Shastry spin model \cite{FukKaw2}.
These observations motivate us to consider a Jastrow-ansatz 
state as an eigenstate for the twisted {\it t-J} model,
\begin{eqnarray}
|\Psi\rangle =\sum_{x_1<\cdots<x_M}&&\sum_{y_1<\cdots <y_Q}
\psi(\{x_\alpha\},\{y_l\};J_s,J_h) \nonumber\\
&&\times\prod_{\alpha}b_{x_\alpha}^\dagger
\prod_lh_{y_l}^\dagger|P\rangle,
\label{EigSta}
\end{eqnarray}
where the wave function $\psi$ is defined by
\begin{eqnarray}
\psi&&(\{x_\alpha\},\{y_l\};J_s,J_h)=
\prod_\alpha z^{J_sx_\alpha}\prod_l z^{J_hy_l}\nonumber\\
&&\times\prod_{\alpha<\beta}d(x_\alpha-x_\beta)^2
\prod_{l<m}d(y_l-y_m)
\prod_{\alpha,l}d(x_\alpha-y_l).
\label{Jas}
\end{eqnarray}
Here, $d(n)=\sin(\pi n/N)$, and we have introduced
\begin{equation}
z=e^{2\pi i/N}.
\end{equation}
Spin and charge currents 
are defined by $J_s=Q/2$ mod 1 and $J_c=(Q+M-1)/2$ mod 1.
Note that the wave function (\ref{Jas}) is the gauge-transformed one
whose basis  is constructed by the periodic
operators $b_{x_\alpha+N}^\dagger=b_{x_\alpha}^\dagger$ 
and $h_{y_l+N}^\dagger=b_{y_l}^\dagger$
(see (\ref{EigSta})). 
The original wave function, based on the twisted operators, is 
obtained via the transformation (\ref{GauTra}),
\begin{eqnarray}
\psi_{\rm org}&&(\{x_\alpha\},\{y_l\};J_s,J_h)\nonumber\\
&&=\prod_\alpha z^{\phi_sx_\alpha}\prod_l z^{\phi_\uparrow y_l}
\cdot\psi(\{x_\alpha\},\{y_l\};J_s,J_h).
\end{eqnarray}
We will show below that the wave function is indeed an eigenstate of 
the present Hamiltonian.
For this purpose, we first calculate the actions of hopping terms
on the wave function (\ref{Jas}) in several steps,
following similar calculations to 
Refs.\cite{KurYok} and \cite{WLC} except that  hopping 
terms are now given by a complicated form  $J_{\phi}$
depending on  $\phi_\sigma$ in a nontrivial way.

\subsection{Action of $T_\uparrow(\phi_\uparrow)$}

The action of $T_\uparrow$ on the wave function is 
\begin{equation}
\frac{T_\uparrow(\phi_\uparrow)\psi}{\psi}=
\sum_{n=1}^{N-1}J_{\phi_\uparrow}(n)z^{J_hn}
\sum_l\prod_{m(\ne l)}F_{lm}^{(n)}\prod_\alpha F_{l\alpha}^{(n)},
\label{Tup}
\end{equation}
where
\begin{equation}
F_{lm}^{(n)}=\cos\frac{\pi n}{N}
+\sin\frac{\pi n}{N}
\cot\Theta_{lm}
\label{Flm}
\end{equation}
with
\begin{equation}
\Theta_{lm}=\frac{\pi(y_l-y_m)}{N}.
\label{Thelm}
\end{equation}
Here we have introduced the same notations as those in \cite{KurYok}.

\subsection{Action of $T_\downarrow(\phi_\downarrow)$}

Note that $T_\downarrow$ exchanges pairs of $\{x_\alpha\}$ and
$\{y_l\}$, and is difficult to treat directly\cite{KurYok,WLC}.
However, if we introduce a unitary operator
$U$ generating the  $\pi$-rotation around
$x$-axis; $U\equiv\prod_je^{\pi i(S_j^++S_j^-)/2}$,
$T_\downarrow$ can be expressed by 
\begin{equation}
T_\downarrow(\phi_\downarrow) =
\sum_{i\ne j}J_{\phi_\downarrow}(i-j)
U^\dagger h_j^\dagger h_iU.
\end{equation} 
Furthermore, from the identity proved by Wang-Liu-Coleman\cite{WLC},
\begin{eqnarray}
\prod_{\alpha<\beta}&&d^2(x_\alpha-x_\beta)
\prod_{\alpha,l}d(x_\alpha-y_l)=\nonumber\\
&&A(M,Q)\prod_lz^{\frac{N}{2}y_l}
\prod_{a<b}d^2(u_a-u_b)
\prod_{a,l}d(u_a-y_l),
\end{eqnarray}
where $A$ is a constant independent of the coordinate,
we find a generalized Wang-Liu-Coleman's theorem including the
twist angle,
\begin{eqnarray}
&&\frac{T_\downarrow(\phi_\downarrow)\psi(\{x\},\{y\};J_s,J_h)}
{\psi(\{x\},\{y\};J_s,J_h)}=\nonumber\\
&&~~~
\frac{T_\uparrow(\phi_\downarrow)
\psi(\{u\},\{y\};N-J_s,J_h-J_s+\frac{N}{2})}
{\psi(\{u\},\{y\};N-J_s,J_h-J_s+\frac{N}{2})}.
\end{eqnarray}
Thus, the action of $T_\downarrow$ can be evaluated in a similar way
to $T_\uparrow$ as 
\begin{equation}
\frac{T_\downarrow(\phi_\downarrow)\psi}{\psi}=
\sum_{n=1}^{N-1}J_{\phi_\downarrow}(n)z^{J_{\downarrow}}
\sum_l\prod_{m(\ne l)}F_{lm}^{(n)}\prod_a F_{la}^{(n)},
\label{Tdow}
\end{equation}
where $J_{\downarrow}=J_h-J_s+N/2$ and $F_{la}^{(n)}$ 
is defined as eq.(\ref{Flm}), 
replacing $y_m$ by $u_a$ in eq.(\ref{Thelm}).

\subsection{Action of $T_s(\phi_s)$}

The action of $T_s$ on the wave function is 
\begin{equation}
\frac{T_s(\phi_s)\psi}{\psi}=
\sum_{n=1}^{N-1}J_{\phi_s}(n)z^{J_sn}
\sum_\alpha\prod_{\beta(\ne\alpha)}B_{\alpha\beta}^{(n)}
\prod_lF_{\alpha l}^{(n)},
\label{Ts}
\end{equation}
where
\begin{equation}
B_{\alpha\beta}^{(n)}=1-g_{\alpha\beta}^{(n)}
\end{equation}
with
\begin{equation}
g_{\alpha\beta}^{(n)}\equiv 
\frac{(1-z^n)z_\alpha^2+(1-z^{-n})z_\beta^2}{(z_\alpha-z_\beta)^2}.
\end{equation}
Here we denote $z_\alpha\equiv z^{x_\alpha}$.

\subsection{Eigenvalues}
There are various many-body terms in eqs.(\ref{Tup}), (\ref{Tdow})
and (\ref{Ts}). 
However, more than three-body terms turn out to
vanish similarly to the case 
of the periodic model. Key identities to 
prove this \cite{Hal,FukKaw2} are
\begin{eqnarray}
S_{st}^\phi (J)&=&\frac{1}{4}
\sum_{n=1}^{N-1}J_\phi(n)z^{Jn}(1-z^n)^s(1-z^{-n})^t\nonumber\\
&=&\left\{\begin{array}{ll}
(-)^s& {\rm for}\quad s+t=2\\
(-)^s(J+\phi-\frac{N}{2})-\frac{1}{2}\quad& {\rm for}\quad s+t=1\\
2\varepsilon(J+\phi)+\frac{1}{3}(N^2-1)&{\rm for}\quad s=t=0 \\
0&{\rm for}\quad{\rm others},
\end{array}\right.
\label{SIde}
\end{eqnarray}
where $J$ is an integer $0<J<N$, $s$ and $t$ are non-negative integers
in the region $0\le s+t\le\min(J+\phi,N-J-\phi)$.
The function $\varepsilon (k)$ for any rational $k$ is defined by
\begin{equation}
\varepsilon(k)=
\left(2[k]-N+1\right)k-[k]\left([k]+1\right),
\label{SinParEne}
\end{equation}
where $[a]$ denotes the Gauss symbol, denoting the maximum integer
which does not exceed $a$.
It is quite characteristic in twisted $1/r^2$ models that 
such a function including the Gauss symbol 
enters the expression.
We will see later that
$\varepsilon(k)$ is nothing but a single-particle energy
as a function  of the momentum $k$. 

Substituting the above results to eqs.(\ref{Tup}), (\ref{Tdow})
and (\ref{Ts}), we have
\begin{equation}
\frac{(T_s+T_\uparrow+T_\downarrow)\psi}{\psi}=
W_0+W_2+W_3.
\end{equation}
First, $W_0$ is the constant term
\begin{eqnarray}
W_0&=&\frac{1}{3}M(N^2-1)+\frac{2}{3}M(M^2-1)+\frac{1}{2}QM(2M-1)\nonumber\\
&&+\frac{2}{3}Q(N^2-1)+\frac{1}{3}Q(Q^2-1)+\frac{1}{2}Q^2(N-Q)\nonumber\\
&&+2M\widetilde\varepsilon(J_s+\phi_s)
+2Q\widetilde\varepsilon(J_h+\phi_\uparrow)\nonumber\\
&&+2Q\left(
\widetilde\varepsilon(J_{\downarrow}+\phi_\downarrow)
+\frac{N^2}{4}\right),
\end{eqnarray}
where $\widetilde\varepsilon(J+\phi)$ is defined by
\begin{eqnarray}
&&\widetilde\varepsilon(J+\phi)\nonumber\\
&&\quad=\left\{
\begin{array}{ll}\varepsilon(J+\phi),&~\mbox{\rm for~ integer}~J,\\
\frac{1}{2}\bigl(\varepsilon(J+\phi-\frac{1}{2})&\\
\quad+\varepsilon(J+\phi+\frac{1}{2})-\frac{1}{2}\bigr),&
~\mbox{\rm for~  half-integer}~J.
\end{array}\right.
\end{eqnarray}
Next, $W_2$ is the two-body term, evaluated as
\begin{eqnarray}
W_2&=&-2i\left(J_s+\phi_s-\frac{N}{2}\right)
\sum_{\alpha,l}\cot\Theta_{\alpha l}\nonumber\\
&&-2i\left(J_h+\phi_\uparrow-\frac{N}{2}\right)
\sum_{\alpha,l}\cot\Theta_{l\alpha}\nonumber\\
&&-2i\left(J_h-J_s+\phi_\downarrow\right)
\sum_{a,l}\cot\Theta_{la}\nonumber\\
&&-\sum_{\alpha\ne\beta}J_0(x_\alpha-x_\beta)\nonumber\\
&=&-\sum_{\alpha\ne\beta}J_0(x_\alpha-x_\beta),
\label{W2}
\end{eqnarray}
which should be canceled out by one of the interaction terms 
in eq.(\ref{IntHam}).
Finally, the three-body term $W_3$ is
\begin{eqnarray}
W_3&=&\frac{1}{2}
\sum_{\alpha\ne\beta,l}
\cot\Theta_{\alpha l}\cot\Theta_{\beta l}
+\frac{1}{2}
\sum_{l\ne m,a}
\cot\Theta_{la}\cot\Theta_{ma}\nonumber\\
&-&\frac{1}{2}
\sum_{a\ne b,l}
\cot\Theta_{al}\cot\Theta_{bl}\nonumber\\
&=&-\sum_{\alpha, l}J_0(x_\alpha-y_l)\nonumber\\
&&+\frac{1}{2}Q(M+1)+\frac{1}{6}Q(N^2-1)\nonumber\\
&&-\frac{1}{2}Q^2(N-M)+\frac{1}{3}Q(Q^2-1).
\label{W3}
\end{eqnarray}
The three-body term thus reduces to the 
two-body term, and is also canceled
out by the other interaction terms in eq.(\ref{IntHam}).
Collecting constant terms, we finally 
end up with the eigenvalue of the Hamiltonian
\begin{equation}
 E_{\rm t}=\left(\frac{\pi}{N}\right)^2(e+E) 
\label{TotEng}
\end{equation}
with
\begin{eqnarray}
e&=&\frac{1}{3}Q(N^2-1),\nonumber\\
E&=&\frac{2}{3}M(M^2-1)\nonumber\\
&+&\frac{2}{3}Q(Q^2-1)+\frac{1}{2}Q(M+Q)(2M-Q)\nonumber\\
&+&2M\widetilde\varepsilon
(J_s+\phi_\uparrow-\phi_\downarrow)\nonumber\\
&+&2Q\widetilde\varepsilon(J_h+\phi_\uparrow)\nonumber\\
&+&2Q\left(
\widetilde\varepsilon(J_{\downarrow}+\phi_\downarrow)+\frac{N^2}{4}
\right).
\label{exactene}
\end{eqnarray}
Here we have divided the energy into two parts
$e$ and $E$ for later convenience, because $E$ can be
reproduced by the asymptotic Bethe ansatz (see eq.(\ref{eee})), 
whereas $e$ simply corresponds to the chemical-potential term.
The currents are subject to the constraints
\begin{eqnarray}
&&\frac{Q}{2}+M-1\le J_s+\phi_\uparrow-\phi_\downarrow\le
N-\left(\frac{Q}{2}+M-1\right),\nonumber\\
&&\frac{1}{2}(Q+M-1)\le J_h+\phi_\uparrow\le
N-\frac{1}{2}(Q+M-1),\nonumber\\
&&-\frac{M+1}{2}\le J_h-J_s+\phi_\downarrow\le \frac{M+1}{2},
\end{eqnarray}
in order for the formula (\ref{SIde}) to be applicable.
Note that when $\phi_\uparrow=\phi_\downarrow=0$, 
$\widetilde\varepsilon(J)$ reduces to the 
well-known form $\widetilde\varepsilon(J)=J(J-N)$,
and the spectrum obtained here coincides with 
that found in \cite{WLC}.

The expression (\ref{exactene}) gives the exact ground state
energy with twisted boundary conditions. However,
in order to trace the flow of the 
initial ground state correctly, we need more detailed information 
to specify the states, which may be supplied by the
motif picture described below\cite{HHTBP}.


\section{Spectral Flow }

\subsection{Asymptotic Bethe Ansatz}
Before discussing the spectral flow as a function of 
the twist angle, we first show that the above exact spectrum can 
be reproduced by the asymptotic Bethe ansatz (ABA), 
which naturally leads us to introduce the 
motif picture. To obtain the ABA equation for the 
supersymmetric {\it t-J} model, we recall that 
the present model is reduced to the Haldane-Shastry
model at $Q=0$, and the spectral flow in this case 
\cite{FukKaw2,FukKaw1} can be described by the ABA equation,
\begin{equation}
\widetilde k_\mu=I_\mu+\phi_s+\frac{1}{2}\sum_{\nu=1}^{M}
\sgn(\widetilde k_\mu-\widetilde k_\nu).
\label{abahs}
\end{equation}
We then have a simple relation
$\tildk_\mu=k_\mu+\phi_s$, in terms of the rapidities $k$'s 
defined at $\phi_s=0$. Namely, the sole effect due to 
$\phi_s$ is to shift $k$'s uniformly.
To observe how the spectral flow occurs, 
it is convenient to introduce the description 
by the motif\cite{HHTBP}, which is briefly outlined 
here in the case of half filling. 
The essential point of the motif in this case is that 
one can describe the effect of two-body phase shifts in 
(\ref{abahs}) by arranging 0 and 1
for a given configuration of rapidities.
For example, the ground state of the Haldane-Shastry 
model is a singlet, denoted by the motif 
\begin{equation}
010\cdots 1010101\cdots 010,
\label{motifs}
\end{equation}
which means that the spacing of the occupied momenta,
denoted by 1's, is enlarged twice as large as that for 
free fermions. From this motif we can identify the 
statistical interaction $g=2$
($g=1$ corresponds to the free fermion case). 
 The spectral flow of this 
state can be described by the flow of the motif,
$010\cdots010\rightarrow001\cdots101\rightarrow010\cdots010$,
where the arrow means $\delta\phi=1$,
or in other words the ring is threaded by a unit flux.
Thus the period of the ground state turns out to be 2, 
which directly reflects the statistical interaction 
$g=2$ of exclusion statistics\cite{HalFES}.

Keeping the above results in mind, we 
now show how the ABA equation for twisted boundaries 
is generalized when holes are doped into the Haldane-Shastry model.
To this end, we exploit the idea of the nested Bethe
ansatz with graded symmetry (supersymmetry) \cite{NBA}, which
leads to the Bethe equation 
for the supersymmetric {\it t-J} model 
with twisted boundary conditions,
\begin{eqnarray}
&&\widetilde k_\mu=I^{(1)}_\mu+\phi_\uparrow-\phi_\downarrow\nonumber\\
&&\qquad\quad-\frac{1}{2}\sum_{i=1}^Q
\sgn(\widetilde k_\mu-\widetilde m_i)
+\frac{1}{2}\sum_{\nu=1}^{M+Q}\sgn
(\widetilde k_\mu-\widetilde k_\nu),
\label{ABAspin}\\
&&I^{(2)}_i+\phi_\downarrow=\frac{1}{2}\sum_{\nu=1}^{M+Q}\sgn
(\widetilde m_i-\widetilde k_\nu),
\label{ABAhole}
\end{eqnarray}
where $\mu=1,2,\cdots,M+Q$ and $i=1,2,\cdots,Q$.
The newly introduced rapidity $\widetilde m_i$
is concerned with the charge degrees of freedom.
The energy $E$ in eq.(\ref{TotEng}) is given by
\begin{equation}
E=2\sum_{\mu=1}^{M+Q}\varepsilon(\tildk_\mu),
\label{eee}
\end{equation}
where the single-particle energy $\varepsilon(\tildk)$ is defined by 
eq.(\ref{SinParEne})\cite{dispersion}.
 Note that the twist 
angles, $ \phi_\uparrow$ and $\phi_\downarrow$, 
have been introduced in the above equations 
so as to be consistent with  a nested 
procedure in the Bethe ansatz \cite{NBA}.
Both of the rapidities $\tildk$ and $\tildm$ are
defined in the range $0\le\tildk,\tildm<N$,
which include the effect of the finite twist angle. 
The corresponding quantum numbers $I^{(1)}_\mu$ 
and $I^{(2)}_i$ should satisfy
 $I^{(1)}_\mu=M/2$ mod 1 and $I^{(2)}_i=(M+Q)/2$
mod 1.

Here we should mention the precise meaning of eq.(\ref{ABAhole})
which has been formally deduced by the nested Bethe ansatz,
because it does not seem to
hold for a fractional value of $\phi_{\downarrow}$ at a first glance. 
This subtle problem comes from the phase shift with
the step-wise sgn($k$) function.
We briefly summarize how to treat this
equation correctly.
Consider, for example, the following configuration at 
$\phi_\uparrow=\phi_\downarrow=0$,
\begin{equation}
\cdots<k_{\mu-1}<m_i<k_\mu<\cdots.
\end{equation}
If we put $\phi_\downarrow=1$, it is seen from 
eq.(\ref{ABAhole}) that 
$I^{(2)}_i$ changes into $I^{(2)}_i+1$ and the above configuration
should be changed to
\begin{equation}
\cdots<k_{\mu-1}<k_\mu<m_i\cdots.
\end{equation}
Namely, $m_i$ exchanges the position with $k_\mu$ 
sitting on its right. Therefore,
fractional $\phi_\downarrow$ between 0 and 1 
should interpolate these two configurations smoothly.
This is naturally realized if we introduce an infinitesimal
width $\eta$ in the step-wise phase shift and
then take the limit of $\eta \rightarrow 0$. 
Based on this observation, we can 
separate the l.h.s. of eq.(\ref{ABAhole}) as 
$I^{(2)}_i+\phi_\downarrow\equiv(I^{(2)}_i+[\phi_\downarrow])
+(\phi_\downarrow-[\phi_\downarrow])$, and the fractional part 
$\phi_\downarrow-[\phi_\downarrow]$ 
can be  absorbed into  
$\sgn(\tildm_i-\tildk_\mu)\equiv\sgn(0)$ in the case of $\tildm_i=\tildk_\mu$
for $0<\phi_\downarrow<1$.
Namely, the fractional portion of $\phi_\downarrow$ 
can be incorporated into sgn(0) by 
taking the appropriate limit mentioned above\cite{interpret}.
As a consequence, we have from eq.(\ref{ABAspin})  
\begin{equation}
\frac{1}{2}\sgn(\tildm_i-\tildk_\mu)=\phi_\downarrow-\frac{1}{2}
\end{equation}
for $0<\phi_\downarrow<1$.
Note that the initial order $m_i<k_\mu$ at
$\phi_\downarrow=0$ indeed changes into 
$k_\mu<m_i$ at $\phi_\downarrow=1$.
Substituting the above equation into eq.(\ref{ABAspin}),
we end up with the simple results,
\begin{equation}
\tildk_\mu=\left\{
\begin{array}{ll}
k_\mu+\phi_\uparrow, \quad&
{\rm for}\cdots<m_i<k_\mu<\cdots ,\\
k_\mu+\phi_\uparrow-\phi_\downarrow ,\quad&
{\rm for}\cdots<k_{\mu-1}<k_\mu<\cdots ,
\end{array}\right.
\label{ExpK}
\end{equation}
for $0\le\phi_\downarrow<1$.
This equation implies that $\tildk$ with (without) $\tildm$ 
in its left neighbor describes the  charge (spin) 
degrees of freedom\cite{Com}.
One can easily confirm that the above ABA equations 
indeed reproduce the exact energy (\ref{exactene}) obtained in 
sect.III by choosing the suitable quantum numbers $I^{(1)}_\mu$
and $I^{(2)}_i$.

\subsection{Spectral Flow in Metallic Phase }

\begin{figure}[h]
\epsfxsize=8cm 
\centerline{\epsfbox{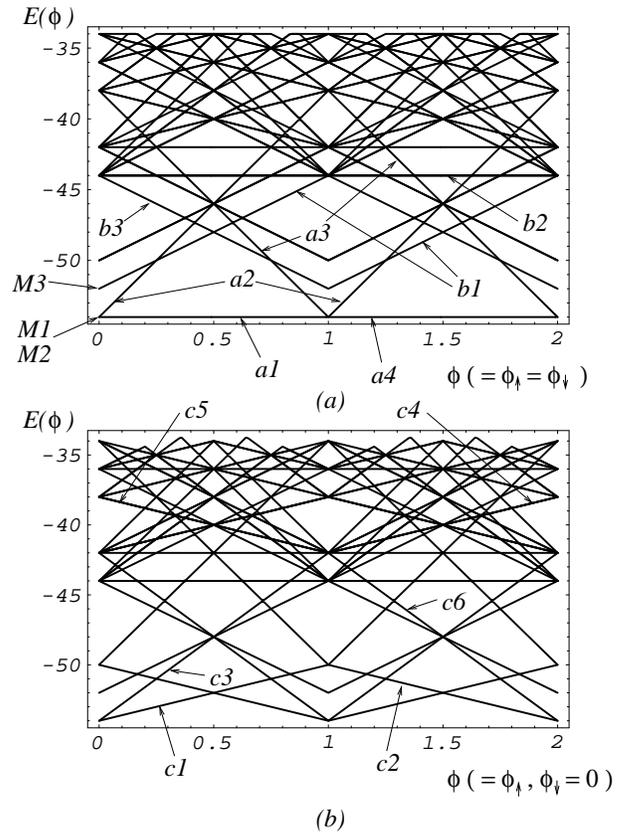}} 
\vspace{0.5cm}
\centerline{
\caption{Exact spectral flow for the system with
$N=6$, $M=2$ and $Q=2$: $\phi=\phi_\uparrow=\phi_\downarrow$ for (a)
and $\phi=\phi_\uparrow, \phi_\downarrow=0$ for (b).
Lower 40 levels are shown. 
In order to change the value of $\phi =p/q$,
integer $p$ is varied consecutively with $q=100$ being fixed.
We have drawn the diagram up to $\phi=2$ for simplicity.
When we discuss the flow beyond $\phi=2$, 
we can use the periodicity of the full spectrum, for example,
$\phi=0,1,2$ should read $\phi=2,3,4$.
Note that the spectral flow exhibits singular cusps, 
which come from the special dispersion relation 
(26) for the present model. See also [6] for detail.
}
\label{fig:fig1}
}
\end{figure}

Having noticed that the exact solution obtained 
in sec.III can be reproduced by the ABA, we are now 
ready to discuss the spectral flow by comparing the 
numerical diagonalization results with the motif picture 
in the ABA.  The numerical results for small systems can be 
complementary to the exact results in sect.III, because 
the analytical method there can supply only a 
special series of the eigenstates.
In Fig.1, we have shown the exact spectral flow
calculated numerically for two holes in the system 
with 6 sites (i.e. 4 electrons). 
In this system, the ground state is doubly degenerate 
concerning the spin degrees of freedom.
For instance, two degenerate 
ground states in Fig.\ref{fig:fig1}
at $\phi_\uparrow=\phi_\downarrow=0$
are classified  by the motifs
\begin{eqnarray}
M1&:&\quad 01\ch 1\ch 1010 ,\nonumber\\
M2&:&\quad 0101\ch 1\ch 10,
\end{eqnarray}
and the first excited state by
\begin{equation}
M3:\quad 01\ch 101\ch 10. 
\end{equation}
In order to express the motif in the doped case,
we have introduced $~\ch~$ in addition to 0 and 1,
where  $~\ch~$ denotes 
the position of $\widetilde m_i$ in eqs.(\ref{ABAspin})
and (\ref{ABAhole}), namely, it specifies a
doped hole in the motif.
When boundaries are twisted, the pattern behavior in the
spectral flow explicitly depends both on the configurations
of $k$'s and $m $'s, because the motions of 
$k$'s  are different from those of $m$'s  as a function
of the twist angle. These behaviors are correctly
described by the flow of $1$ and  $~\ch~$ in the motif.
Since we can change two independent twist angles
in the doped case, we have
examined the spectral flow for two typical cases in
Fig.\ref{fig:fig1}.
One is the case where $\phi_\uparrow=\phi_\downarrow=\phi$,
which will be referred to as Type I.
In this case, the twisted boundary condition is imposed only
on the charge degree of freedom.
The other is the case where $\phi_\uparrow=\phi, \phi_\downarrow=0$,
which will be referred to as Type II.  It is seen from eq.(\ref{ExpK})
that the twisted boundary condition in Type II
is equally imposed both on the charge and
 spin degrees of freedom.
The motif $M1$, for example, behaves
\begin{eqnarray}
&&01\ch 1\ch 1010\rightarrow 0101\ch 1\ch 10\quad{\rm for~Type~I},\nonumber\\
&&01\ch 1\ch 1010\rightarrow 001\ch 1\ch 101\quad{\rm for~Type~II},
\end{eqnarray}
when we increase  the twist angle
 $\delta\phi=1$.
Therefore the corresponding spectral flow reads
\begin{eqnarray}
M1&:& a1\rightarrow a2\rightarrow a3\rightarrow a4~(=a1),\nonumber\\
M2&:& a2\rightarrow a3\rightarrow a1\rightarrow a2,\nonumber\\
M3&:& b1\rightarrow b2\rightarrow b3\rightarrow b1,
\end{eqnarray}
for Type I shown in Fig.1(a), and
\begin{eqnarray}
M1&:& c1\rightarrow c2\rightarrow c3\rightarrow 
c4\rightarrow c5\rightarrow c6\rightarrow c1,\nonumber\\
M2&:& c3\rightarrow c4\rightarrow c5\rightarrow 
c6\rightarrow c1\rightarrow c2\rightarrow c3 ,
\end{eqnarray}
for Type II ($M3$ is the same) shown in Fig.1(b).
Note that $\rightarrow$ means $\delta\phi=1$.
In this way we can trace the natural
spectral flow correctly following the flow of the motif. 
This in turn demonstrates that the spectral flow in the 
supersymmetric {\it t-J} model with $1/r^2$
interaction completely fits in with the motif picture in the ABA
even in the metallic case.
We note that the spectral flows $M1$ and $M2$  for the ground states
(Fig.\ref{fig:fig1}) agree with the exact results obtained in 
sect.III.

Based on these results for small systems, 
we now wish to generalize our statement to more 
generic cases, and deduce 
how the spectral flow behaves when any number 
of holes are doped into
the Haldane-Shastry model with $N$-sites.
We first note that the doped holes energetically prefer to
occupy the positions in the center of the motif.
For example, if we dope one or two holes,
the corresponding motif should be
\begin{eqnarray}
010\cdots 1010\ch 101\cdots 010 ,\nonumber\\
010\cdots 101\ch 1\ch 101\cdots 010.
\label{mhole}
\end{eqnarray}
It is easily seen from these motifs that the period of the flow
becomes macroscopic, i.e.  $N$ 
both for Types I and II. Note  that this is 
also the case for any doping rates away from half filling, 
namely, the period generally becomes $N$ both for
Type I and Type II.
The period $N$ for Type I  (the charge sector)
is naturally understood by the analogy to the free fermion case.
On the other hand, the change of the period 
in Type II is remarkable, because 
it is given by 2 at half filling, reflecting fractional exclusion 
statistics with statistical interaction $g=2$. 
Therefore once holes are doped, characteristic properties of 
ideal exclusion statistics for the spin sector
 are hidden in the spectral flow.
This does not imply, however, that 
the nature of exclusion statistics is spoiled by doping, because 
various characteristic properties such as correlation functions
can be still interpreted in terms of \
exclusion statistics\cite{HaHalTJ,KatKur}.
A reason why we cannot observe exclusion  statistics
in the spectral flow is related to  breaking  of the 
``commensurability'' condition. 
For example, exclusion statistics with $g=2$ is 
commensurate at half filling in the sense that
the corresponding motif for the 
ground state fits in with the underlying lattice
(see eq.(\ref{motifs})).
In contrast, the motif
for the metallic phase, for example, (\ref{mhole}) does not 
meet the commensurability condition, hiding the 
$g=2$ statistics in the flow.
Therefore, we can say that the commensurability condition 
is essential to observe
 the ideal exclusion statistics in the spectral flow.

In this connection, it is instructive 
to think of a specific motif for an excited state 
in the case of $N/3$ hole-doping, 
which has a uniform configuration such that
\begin{equation}
01\ch 101\ch 10\cdots 01\ch 10 .
\end{equation}
This motif satisfies the commensurability 
condition, and the resultant period of the state turns out 
to be 3. Although this kind of a special excitation
may not be interesting for the present model, we wish to note that 
the same motif  describes
the ``ground state'' for the SU(3) Haldane-Shastry spin chain.
Therefore, we can predict that the ground state of the
latter model, or more generally, SU($\nu$) spin model can have a 
period $\nu$ in the spectral flow.

The above characteristic properties for 
the $1/r^2$  model are to be compared with ordinary 
interacting electron systems, for which
the ideal statistics may not be realized 
due to perturbations which deviate the system 
from the idealized situation.
In such cases, it is difficult to observe 
exclusion statistics in the spectral flow even if the 
commensurability conditions are met,
because irrelevant perturbations may
dominate such properties.

\section{Summary}

We have investigated the spectral flow in 
the supersymmetric {\it t-J} model with 
$1/r^2$ interaction by analyzing the exact spectrum 
with twisted boundary conditions.
At half filling, exclusion statistics 
with statistical interaction $g=2$ 
can be observed in the period of the spectral flow.
Away from half filling, two kinds of spectral flows for the 
charge and spin sectors appear. It has been found 
that the nature
of the spectral flows nicely fits in with the motif picture 
even in the metallic phase.  We have also addressed the question
whether one can observe fractional exclusion statistics
explicitly in the spectral flow in the metallic phase. 
By analyzing the exact spectrum in terms of 
the motif picture, we found that 
the period becomes $N$
once any number of holes are doped, hiding 
characteristic properties of exclusion statistics with $g=2$
for the spin sector.
This drastic change has been shown to be related
to breaking of the commensurability condition.
Therefore, even in the idealized model for exclusion statistics,
which is free from any irrelevant perturbations,
it is crucial to satisfy the commensurability condition 
to observe the statistical interaction in the 
period of the spectral flow.

In this paper, we have been concerned with the 
cases of rational twist angles in unit of $2\pi$.
We think that the present conclusion can be applied
even for irrational cases, although we could not
yet find the exact solution for such cases.

\acknowledgements{
This work is supported by the Grant-in-Aid from the Ministry of
Education, Science and Culture, Japan.}


{\it Note added in proof}

We have learned that the spectral flow for the $XXZ$ model with
short-range interaction was studied in 
 F. C. Alcaraz, M. N. Barber and M. T. Batchelar, 
Ann. Phys. {\bf 182} (1988) 280 .


\end{multicols}

\begin{references}
\bibitem{dispersion} In the case of the periodic model, 
the momentum $k$ is given by an integer and
thus the function (\ref{SinParEne})
is reduced to the quadratic dispersion relation
$\varepsilon(k)=k(k-N)$ known for $1/r^2$ models.
A remarkable point is that    
the dispersion relation (\ref{SinParEne}) for fractional cases
is now essentially {\it linear} in $k$.
It was for the first time found in \cite{FukKaw2}.
\bibitem{interpret} 
This kind of the interpretation is also necessary 
to reproduce the free electron spectrum with twisted boundary
conditions by using the Bethe equation of
the Hubbard model\cite{ShaSut} 
in its zero interaction limit.
\end{references}
\end{document}